\documentclass[conference,11pt]{IEEEtran} 

\usepackage{framed}
\usepackage{listings}
\usepackage{latexsym}
\usepackage{verbatim}
\usepackage[pdftex]{graphicx}
\usepackage{multirow}
\usepackage[algoruled,linesnumbered,lined]{algorithm2e}
\usepackage[hyphens]{url}
\usepackage{array}
\usepackage{flushend}
\usepackage{amssymb}
\usepackage{amsmath}
\usepackage{color}
\usepackage{enumerate}
\usepackage{threeparttable}
\usepackage[font=footnotesize,caption=false]{subfig}
\usepackage[normalem]{ulem}
\usepackage{amsmath}
\usepackage{mathabx}
\usepackage{etoolbox}
\usepackage[inline]{enumitem}

\newcommand{\ignore}[2]{\hspace{0in}#2}

\usepackage[text={18.4cm,23.95cm}]{geometry} 

\usepackage{hyperref}
\hypersetup{
    bookmarks=true,         
    unicode=false,          
    pdftoolbar=true,        
    pdfmenubar=true,        
    pdffitwindow=false,     
    pdftitle={A Study on the Impact of Wind Generation on the Stability of Electromechanical Oscillations},    
    pdfauthor={Charalambos Konstantinou},     
    pdfsubject={Wind Stability},   
    pdfcreator={Dr. Harish Kumar},   
    pdfproducer={Charalambos},  
    pdfkeywords={wind} {stability} {oscillations}, 
    pdfnewwindow=true,      
    colorlinks=false,       
    linkcolor=red,          
    citecolor=green,        
    filecolor=magenta,      
    urlcolor=cyan           
}

\pdfoutput=1
\begin{document}

\title{A Study on the Impact of Wind Generation on the Stability of Electromechanical Oscillations}

\author{
\IEEEauthorblockN{\textbf{Charalambos Konstantinou}}
\IEEEauthorblockA{Electrical and Computer Engineering\\New York University Polytechnic School of Engineering\\E--mail: ckonstantinou@nyu.edu}
}

\maketitle

\thispagestyle{plain}

\begin{abstract}

Wind is becoming an increasingly significant source of energy in modern power generation. Amongst existing technologies, Variable Speed Wind Turbines (VSWT) equipped with Double Fed Induction Generators (DFIG) is widely deployed. Consequently, power systems are now experiencing newer power flow patterns and operating conditions. This paper investigates the impact of a DFIG based Wind Farm (WF) on the stability of electromechanical oscillations. This is achieved by performing modal analysis to evaluate the stability of a two-area power network when subjected to different wind penetration levels and different geographical installed locations. The approach via eigenvalues analysis involves the design of voltage and Supplementary Damping Controllers (SDCs) that contribute to network damping. The effect of Power System Stabilizer (PSS) is also examined for several network conditions. Simulations demonstrate a damping improvement up to 933\% when the control systems are activated and the system operates with 25\% wind integration. 

\end{abstract}

\pagestyle{plain}
\section{Introduction}

\begin{figure}[t]
  \centering
  \includegraphics[width=3.5in]{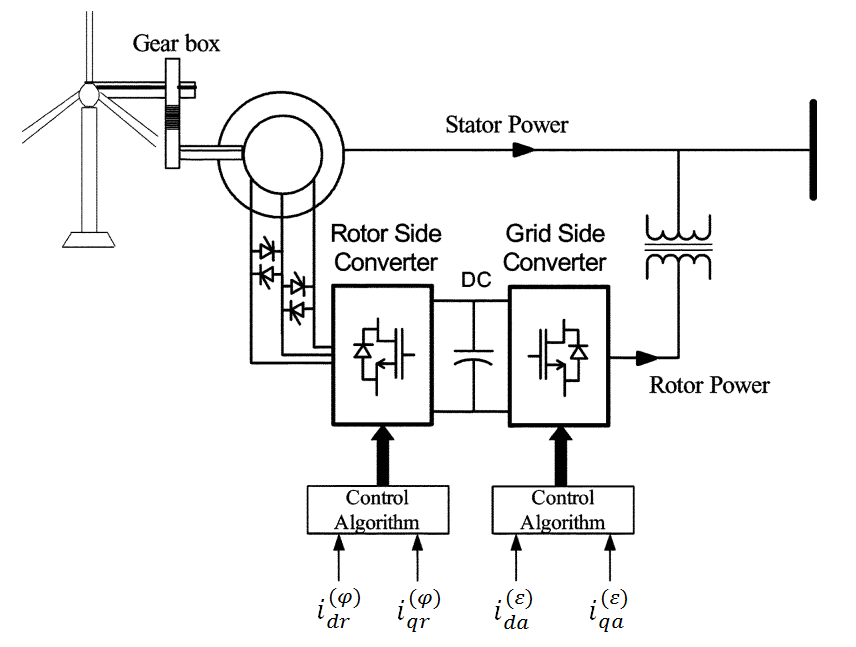}
  \caption{DFIG based WT generator \cite{ref2}.}
  \label{f:fig1_label}
\end{figure}

Today, DFIGs are the most popular wind power generators in the wind power industry. The use of this machine has certain advantages, such as variable speed, active-reactive power control (by using converters) and the economic cost \cite{ref1}. The layout of a DFIG Wind Turbine (WT) is shown in Fig. \ref{f:fig1_label} \cite{ref2}. Between the rotor and the grid there are two linked converters in a back to back pattern. The only requirement for the power electronic converters is the percentage that must be rated in order to handle a fraction of the total power (rotor power). In most cases, this percentage is about 30\% of the nominal generator power \cite{ref3}. The two converters are connected through a dc-link capacitor which, as an energy storage element, maintains small voltage variations in the dc-link. The control of the torque (or the speed) of the DFIG as well as the power factor at the stator terminals is achieved with the Rotor Side Converter (RSC). The main objective for the Grid Side Converter (GSC) is keeping the dc-link voltage constant despite of the magnitude and direction of the rotor power. The GSC operates at the grid frequency, primarily to produce (leading) or to consume (lagging) reactive power.  Typically, a transformer is connected in the middle of the GSC, or the stator, and the grid.  The working frequency of RSC depends on the wind speed \cite{ref2}. 

\begin{figure}[t]
  \centering
  \includegraphics[width=2.6in,height=1in]{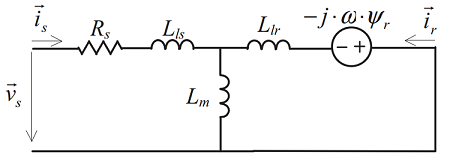}
  \caption{DFIG equivalent circuit \cite{ref1}.}
  \label{f:fig2_label}
\end{figure}

Fig. \ref{f:fig2_label} shows the equivalent model of the turbine according to Park’s equations, referenced to the stator. The circuit contains the speed voltage $\varpi\psi$ and the stator and rotor components \cite{ref1}. 

The authors in \cite{ref3} discuss how wind power and the type of the WT generator change the damping and the frequency of eigenvalues. In particular, the study shows that wind power tends to improve the damping of inter-area oscillations while the impact on intra-area is not significant. In \cite{ref4} it is shown that New Zealand system’s damping performance is not significantly affected by increasing the wind penetration level. An efficient way to increase the damping of electromechanical oscillations using an auxiliary PSS loop installed in DFIG WT generators is presented in \cite{ref5}. Fernandez \textit{et al.} \cite{ref6} add controllers in active and reactive power control loops of DFIG to enhance the damping of oscillations. In addition, synchronous generators' replacement with DFIG based WF in \cite{ref7}, it is shown that does not degrade the two area system modal characteristics. Nevertheless, WF improves the stability of the inter-area mode of the power system. A damping control method for a STATCOM in a series compensated WF for mitigating subsynchronous resonance and damping oscillations is analyzed in \cite{ref8}. In \cite{ref9} non-linear controllers are implemented for improving the damping of intra-area and inter-area oscillations. 

The above references verify the impact of wind generations on electromechanical oscillations. In most cases the goal is to design controllers that can improve network damping. However, the control methods presented in literature can be implemented only for certain systems and do not have universal usage. Moreover, the WFs consist hundreds of WT generators distributed on vast areas. Consequently, it is perplexing to select the feedback signals. Also, conventional PSS methods are practically difficult to implement into the WT generators since their capacity is small and their number is greater compared to the synchronous machines. These differences between conventional power plants and WFs necessitate the design of a controller for DFIG based WFs which has simple theory and can be easily implemented.

The layout of this paper is as follows: It begins by explaining the concepts of RSC, GSC and the dc-link of back-to-back converter. Section \ref{s:modal_analysis} describes the theory behind modal analysis. In Section \ref{s:control} the voltage and damping control strategy is presented. Section \ref{s:system} presents the test system and the study scenarios. Finally in Section \ref{s:results}, simulation results are presented and comments on the results of the system performance are addressed.

\section{Description of the back-to-back converter}\label{s:back_converter}
This section discusses the arrangement of the back-to-back converter. It presents the rotor-side and grid-side current control reference frames and explains the utility of the dc-link.

\subsection{Rotor side converter}

The RSC applies the voltage to the rotor windings of the DFIG and its purpose is to control the rotor currents. Specifically, it controls the developed torque at the shaft of the machine as far as the rotor flux position is oriented in an optimal way with respect to the stator flux.
 
On the rotor perspective, the current is disintegrated in two mechanisms: $i_{dr}^{(\varphi)}$ in phase with the stator flux linkage and $i_{qr}^{(\varphi)}$ in quadrature. The electromagnetic rotation and the machine excitation are decoupled and measured by means of $i_{qr}^{(\varphi)}$ and $i_{dr}^{(\varphi)}$, correspondingly. Fig. \ref{f:fig3_label} (a) shows the association between the rotor current phasor $I_r$, the stator flux linkage $\psi_s$, the current components used by the control system $i_{qr}^{(\varphi)}$ and $i_{dr}^{(\varphi)}$, a synchronous rotating reference frame ${d,q}$ which formulates an angle $(\varphi)$ with the stator flux linkage, and the current components in such reference frame $i_{dr}$ and $i_{qr}$ \cite{ref10}.

The RSC regulates the WT output power and voltage (or reactive power) at the machine's stator terminals by using a torque controller such that the power will follow a turbine power-speed characteristic (tracking the maximum power point). The actual power from the generator terminals ($P_{out}$),  added to the total power losses, is compared with the reference power ($P_{ref}$) obtained from the WT characteristic. In most cases, a Proportional-Integral (PI) regulator is used at the outer control loop to reduce the error (power/rotor speed error). The output of this regulator is $i_{qr}^{(\varphi)}$ i.e. the reference rotor current that must be injected into rotor winding by the RSC. 

\subsection{Grid side converter}

The GSC is designed in such way to regulate the capacitor voltage of the dc bus. In addition, it can generate or absorb reactive power to provide voltage support whenever necessary. 

On the grid side, the current is also disintegrated in two mechanisms: $i_{da}^{(\varepsilon)}$ in phase with the stator voltage and $i_{qa}^{(\varepsilon)}$ in quadrature. The active and reactive power exchanged with the grid are also decoupled and controlled by means of $i_{da}^{(\varepsilon)}$ and $i_{qa}^{(\varepsilon)}$, respectively. In this kind of control systems, the active power output ($P_{out}$) of the GSC is used to manage the dc – link voltage level. 

\begin{figure}[t]
  \centering
  \includegraphics[width=3.5in]{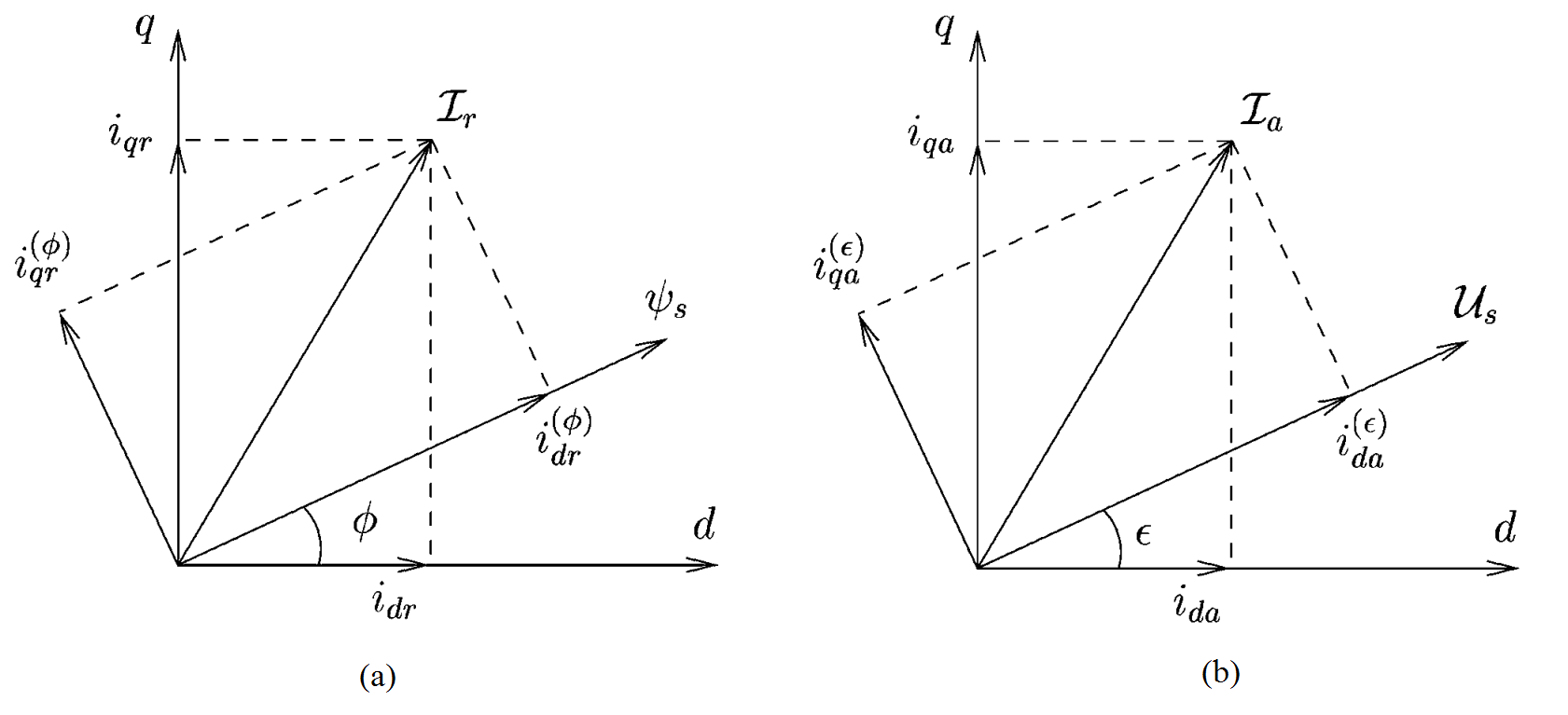}
  \caption{(a) Rotor-side current control reference frame,  (b) Grid-side current control reference frame \cite{ref10}.}
  \label{f:fig3_label}
\end{figure}

Fig. \ref{f:fig3_label} (b) shows the connection between the stator voltage $u_s$, the GSC current $I_a$, the control system current components $i_{da}^{(\varepsilon)}$) and $i_{qa}^{(\varepsilon)}$, a synchronous rotating reference frame ${d,q}$ which formulates an angle $(\varepsilon)$ with the stator voltage, and the current elements in this kind of reference frame $i_{da}$ and $i_{qa}$ \cite{ref10}. 


\subsection{DC-Link}

The dc-link capacitor variations of the voltage are represented by the dc-link model which takes as input the power. Eq. \ref{eq:eq_1} shows the energy stored in the dc capacitor \cite{ref10}:

\begin{equation}
    \label{eq:eq_1}
    W_{dc} = \int P_{dc} dt = \frac{1}{2} \cdot C \cdot V_{dc}^2
\end{equation}

{\setlength{\parindent}{0cm}
where $W_{dc}$ is the stored energy and $P_{dc}$ is the dc-link input power. Eq. \ref{eq:eq_2}} presents the energy and voltage derivatives:

\begin{equation}
\begin{split}
    \label{eq:eq_2}
    \dfrac{dV_{dc}}{dt} = \dfrac{P_{dc}}{C \cdot V_{dc}} \\
    \dfrac{dW_{dc}}{dt} = P_{dc}
\end{split}
\end{equation}

{\setlength{\parindent}{0cm}
The $P_{dc}$ is calculated as $P_{dc} = P_{in} -  P_{c}$. $P_{in}$ is the RSC input power and $P_c$ is the GSC output power. The voltage at the dc-link is constant when $P_{dc} = 0$ and varies with respect to $P_{dc}$.}

The GSC controller monitors the dc-link voltage which essentially determines the exported power by the GSC. For example, if the dc-link voltage increases then the GSC can export more real power by increasing the load angle such that the voltage will reduce back to its nominal value. In simple terms, the dc-link voltage represents the power flow balance between generated and exported energy in the rotor side. In the case where the dc-link capacitor input and output power do not match, then the dc-link voltage will change accordingly \cite{ref14}.

\section{Modal Analysis}\label{s:modal_analysis}
Modal analysis of angle stability is concerned with the characteristic modes determination of a system mode linearized about its  Stable or Unstable Equilibrium Points (SEP, UEP). It primarily comprises computing eigenvalues, eigenvectors and participation factors. Particularly, analysis of these quantities uncovers the involvement of those machines that are losing synchronism, as for example in an inter-area mode oscillation \cite{ref16}. \ignore {While post-disturbance SEPs are generally easy to determine, computation of UEPs has long been considered as a problematic task \cite{ref17} \ignore{ref18}.} Modal identification \ignore{, on the other hand,} relates to the perseverance of characteristic modes from large-disturbance dynamic behavior, that is gained through transient stability simulations.

The non – linear model that defines the dynamics of a power system, includes a set of differential and algebraic equations (Eq. \ref{eq:eq_3} and \ref{eq:eq_4}) \cite{ref21}: 

\begin{equation}
    \label{eq:eq_3}
    \dot{x} = f(x,y)
\end{equation}

\begin{equation}
    \label{eq:eq_4}
    0 = g(x,y)
\end{equation}

{\setlength{\parindent}{0cm}
where x and y are the vectors of state variables and algebraic variables, correspondingly.}

A conventional modal analysis is implemented by linearizing the system equations in Eq. \ref{eq:eq_3} about an equilibrium point, $x_o$, defined by:
\begin{equation}
    \label{eq:eq_5}
    x_o = f(\dot{x_o}) = 0
\end{equation}

Small disturbances around $x_o$ can then be defined by:
\begin{equation}
    \label{eq:eq_6}
    \Delta\dot{x} = A \cdot \Delta x
\end{equation}

{\setlength{\parindent}{0cm}
where $A$ is the system state matrix defined by:}

\begin{equation}
    \label{eq:eq_7}
    A =  \langle A_{ij} \rangle  =  \langle \dfrac{\partial f_i}{\partial x_j}\vert _{x=x_o} \rangle
\end{equation}

The characteristic modes of Eq. \ref{eq:eq_6} are generally formed as:
\begin{equation}
    \label{eq:eq_8}
    \varphi_i \cdot e^{\lambda_i t}
\end{equation}
{\setlength{\parindent}{0cm}
where $\varphi_i$ is the characteristic vector (or right eigenvector) and $ \lambda_i $ is the corresponding eigenvalue. Eigenvalues are solutions of the characteristic equation:}
\begin{equation}
    \label{eq:eq_9}
    det ( A - \lambda I) = 0
\end{equation}

Lyapunov stability theory states that if all the eigenvalues of matrix $A$ have negative real parts, then the equilibrium point is small signal stable \cite{ref21}. If the eigenvalues are complex and located near the imaginary axis of the complex plane, then are characterized as critical. For an eigenvalue $\lambda = \sigma \pm j \omega$, the frequency of the oscillation is given by $\omega / 2\pi$ and the damping ratio $\zeta$ is given by:
\begin{equation}
    \label{eq:eq_10}
   \zeta = - \sigma / \sqrt{\sigma^2 + \omega^2}
\end{equation}

The mode shape related to each critical eigenvalue can be used to identify the oscillation types and therefore the generators' location in the power system \cite{ref21}. \ignore{, ref22} In a power system with multiple generators, the dominant electromechanical oscillation mode often present two clusters relative swing. Furthermore, when the electromechanical system is being low-frequency oscillation, the increment of angular speed and power angle of two generators have the characteristics of sinusoidal oscillation with same frequency and reversed phase. 

\section{Voltage and damping control systems}\label{s:control}

\subsection{Voltage Control}\label{ss:v_control}

\begin{figure}[t]
  \centering
  \includegraphics[width=3.5in]{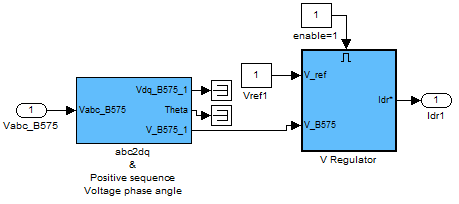}
  \caption{Voltage Controller.}
  \label{f:fig4_label}
\end{figure}

The high penetration level of wind generations affects the power flows and hence the node voltages \cite{ref24}. Due to the fact that WFs replace conventional generators, WFs must contribute to voltage control. As a result, technical regulations for wind power generation are growing at the same rate at its installation \cite{ref25}. The basic approach for the voltage control is to be centralized at the Point of Common Coupling (PCC) that is the voltage bus to which the WF is connected. Fig. \ref{f:fig4_label} illustrates the configuration of the voltage controller. Voltage is transformed from three-phase to direct and quadrature $(dq)$ components. The reference voltage that we want to maintain to the voltage bus is added to $(dq)$ voltage and through a PI controller we get the direct axes current. This current controls the $dq$ voltage of the RSC and hence we are able to regulate the voltage to the desired level.

\subsection{Supplementary damping control}\label{ss:d_control}

\begin{figure}[t]
  \centering
  \includegraphics[width=3.5in]{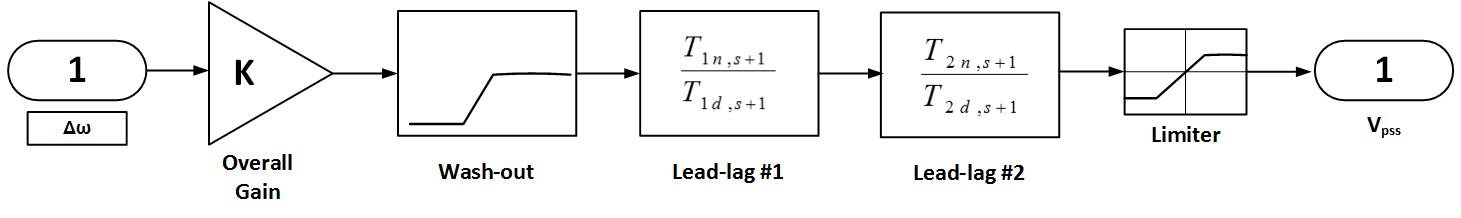}
  \caption{Supplementary Damping Controller.}
  \label{f:fig5_label}
\end{figure}

The unbalanced rotor torque of synchronous generator in power systems has the effect to cause electromechanical oscillations. An effective solution is adding a supplementary damping signal in the active power control loop.  \ignore {In \cite{ref26}, damping controllers based on the Phillips-Heffron model are designed for damping low frequency oscillations. In this paper, the damping mechanism is presented and implemented to DFIG RSC control loop from the perspective of dynamic frequency.} The configuration of the SDC system with the traditional lead-lag compensation structure which contributes to improve the damping ratio of machine’s torque is shown in Fig. \ref{f:fig5_label}. It contains a washout filter, a gain block and two lead-lag blocks. The transfer function of the SDC system is presented in Eq. \ref{eq:eq_11}.

\begin{equation}
    \label{eq:eq_11}
   \small V_{pss} (s) = [K] \cdot \left[ \dfrac{sT_w}{sT_{w} + 1} \right]  \cdot \left[ \dfrac{sT_{1n} + 1}{sT_{1d} + 1} \right] \cdot \left[ \dfrac{sT_{2n} + 1}{sT_{2d} + 1} \right] 
\end{equation}

{\setlength{\parindent}{0cm}
where $K$ is the SDC system gain, $T_w$ is the washout time constant and $T_{1n}$, $T_{1d}$, $T_{2n}$, $T_{2d}$ are the time constants of the lead-lag blocks. The washout time constant prevents steady changes in speed from modifying the field voltage, whereas the time constants of lead-lag blocks provide the necessary phase compensation.}

The design of the SDC system for WTs is based on the PSSs used for conventional generators. The input can be any signal affected by the oscillation. This fact implies the selection of the PCC as measurement point in order to avoid the filtering effect introduced by the transformer which is connected between the grid and the WF. In this study, the rotation rate deviation $\Delta\omega$ of WF at the PCC is used as input signal. The amplification block affects the amplitude frequency characteristic. The role of washout block is filtering the direct current component. The lead-lag block is used to compensate the phase lag caused by the converter. The output signal can be any variable capable of varying the power delivered to the grid. For our case the excitation voltage is used as the output signal. 

The active power transfer function of the excitation controls with SDC system in the current regulator of the RSC control system is given by Eq. \ref{eq:eq_12}:

\begin{equation}
    \label{eq:eq_12}
   G(s) = \dfrac{\Delta P(s)}{\Delta \omega (s)} = C (sI - A)^{-1}B
\end{equation}

\section{Test system and study scenarios}\label{s:system}

\subsection{Test system}\label{ss:test_system}
\begin{figure}[t]
  \centering
  \includegraphics[width=3.5in]{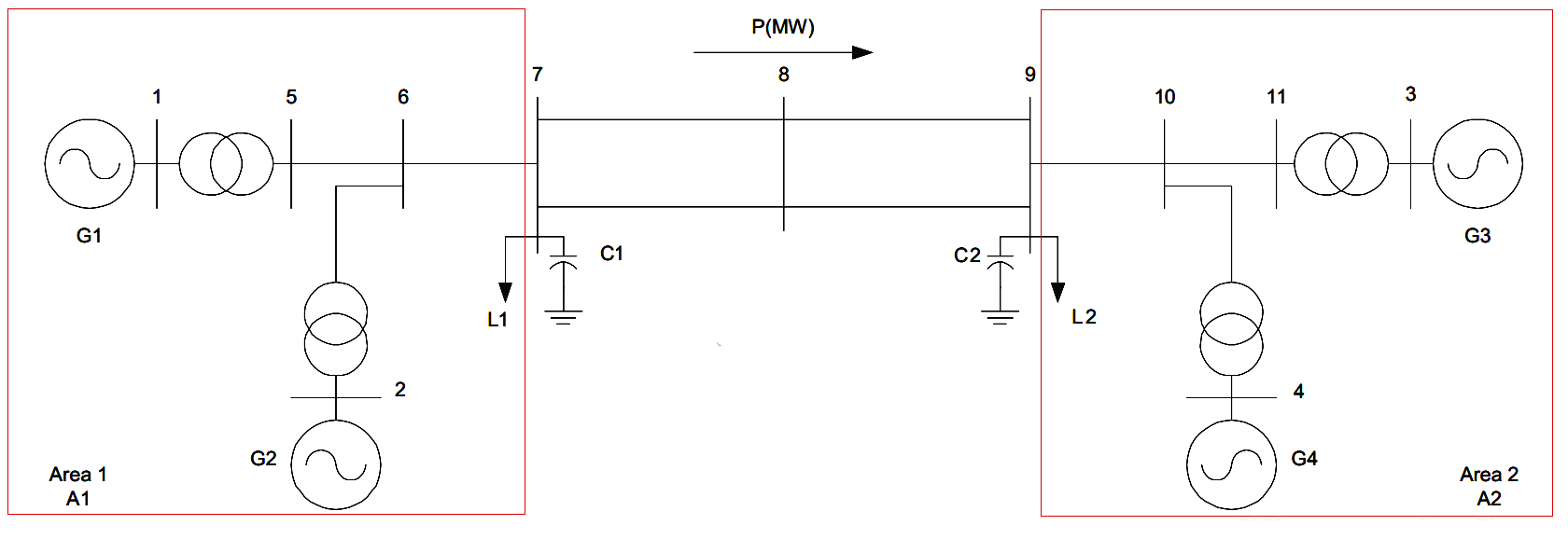}
  \caption{Kundur's two-area system.}
  \label{f:fig6_label}
\end{figure}

To evaluate the impact of a DFIG based WF on the electromechanical oscillations, the two-area power network shown in Fig. \ref{f:fig6_label} \cite{ref21} is examined. The system consists two identical areas connected through a relatively weak tie. Each area includes two synchronous generators with equal power outputs and loads which are modeled as constant impedance (full symmetry). Three electromechanical oscillations are present in this system. Two intra-area modes (i.e. generators' oscillations in a certain area of the network against each other), one in each area, and one inter-area low frequency mode (i.e. generators' oscillations in a certain area of the network against generators in another area of the network). 

\begin{figure}[t]
  \centering
  \includegraphics[width=3.5in]{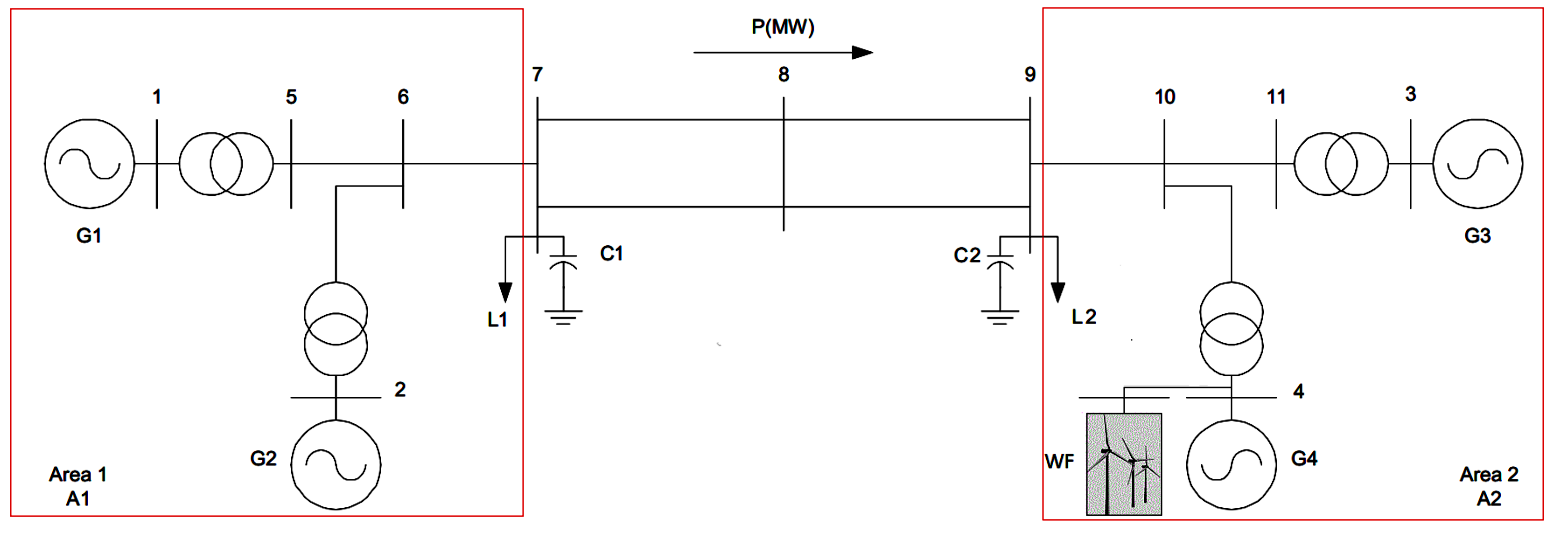}
  \caption{Kundur's two-area system integrated with a WF next to the synchronous generator G4.}
  \label{f:fig7_label}
\end{figure}

\begin{figure}[t]
  \centering
  \includegraphics[width=3.5in]{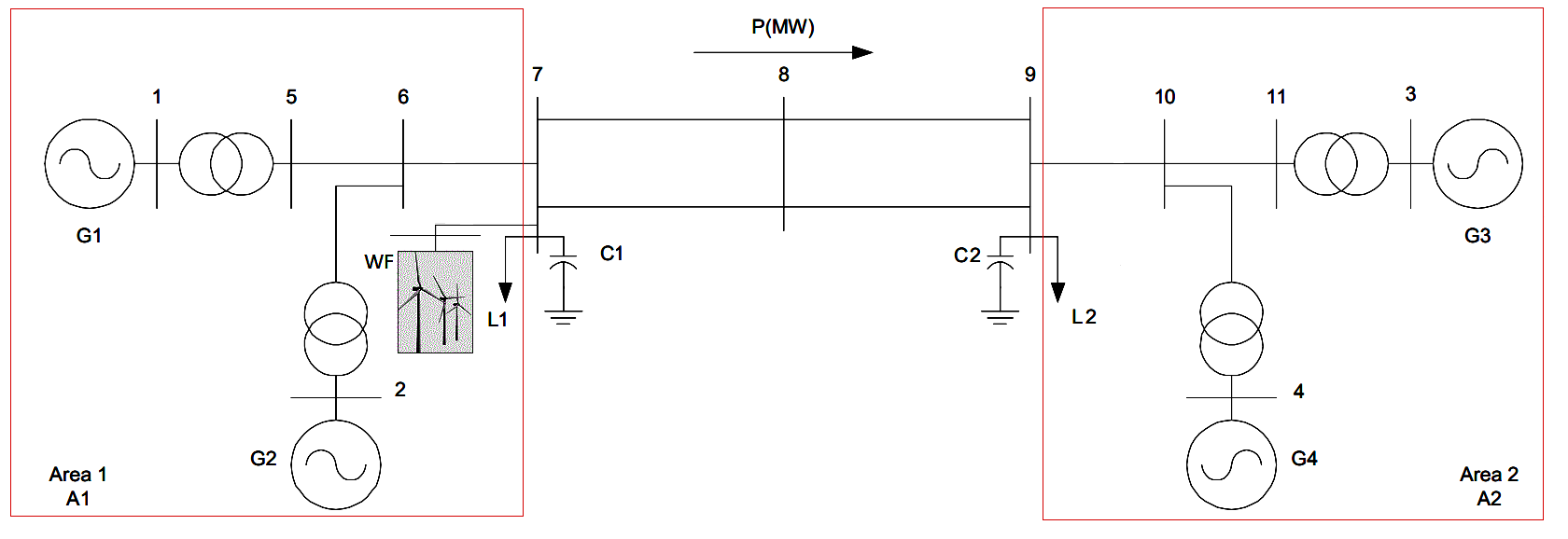}
  \caption{Kundur's two-area system integrated with a WF next to the load L1.}
  \label{f:fig8_label}
\end{figure}

The parameters of the study include different wind penetration levels and different geographical installed locations. The examined level of wind generation is up to 35\% of the parallel synchronous machine capability. Locations constitute the connection of a WF next to the synchronous generator G4 and next to load L1 as shown in Fig. \ref{f:fig7_label} and Fig. \ref{f:fig8_label}, respectively. 

\subsection{Study scenarios}\label{ss:scenarios}

In the case studies, four scenarios are considered. For each case study, the system is simulated with and without PSS for the synchronous conventional generators. Furthermore, a 3-phase fault takes place at the tie-line between the east and west area of the system at \emph{1.0s} and it vanishes \emph{0.2s} later: 

\noindent{\bf Case 1}: Without WF connected to the system (Fig. \ref{f:fig6_label}). 

\noindent{\bf Case 2}: With WF integrated to the system next to the synchronous generator G4 (Fig. \ref{f:fig7_label}). In order to distinctly highlight the fundamental characteristics of WF, no controls are applied to the WF. 

\noindent{\bf Case 3}: With WF integrated to the system next to the load L1 (Fig. \ref{f:fig8_label}). Like the previous case, no controls are applied to the WF.

\noindent{\bf Case 4}: With WF included to the system next to the load L1 (Fig. \ref{f:fig8_label}). In this case, voltage controller and SDC are integrated to the system.

\section{Simulation Results}\label{s:results}
The validation of the presented methods is performed by extracting the eigenvalues of the linearized state-space model using MATLAB/Simulink. The results regarding inter-area and intra-area oscillations are shown in Tables \ref{tab:table1}, \ref{tab:table2}, \ref{tab:table3} and \ref{tab:table4} for the Cases 1, 2, 3 and 4 respectively.

\begin{table}[t]
\centering
\caption{Results of two-area four machine system}
\label{tab:table1}
\renewcommand{\arraystretch}{1}
\begin{tabular}{||c|c|c|c|c||}
\hline
\hline
\multirow{2}{*}{\parbox{1.1cm}{\centering {\bf Case 1}\\}}  & 
\multirow{2}{*}{\parbox{1.4cm}{\centering {{{\bf Eigenvalue}}}}} & 
\multirow{2}{*}{\parbox{1.1cm}{\centering {{{\bf Damping}}}}} & 
\multirow{2}{*}{\parbox{0.8cm}{\centering {{{\bf Freq. \\(Hz)}}}}} & 
\multirow{2}{*}{\parbox{1.2cm}{\centering {{{\bf Oscillation \\ mode}}}}}  \\
{}     &       {}      &    {} &      {}     &       {} \\ \hline
\hline
\multirow{3}{*}{\bf {No PSS}}     &0.108 $\pm$ 4.027i     & \bf{-0.027} & 0.641 & \bf{Inter-area}\\
{}                          &-0.677 $\pm$ 7.048i	&  0.096 & 1.122 & Intra-area1\\
{}                          &-0.669 $\pm$ 7.269i	&  0.092 & 1.157 & Intra-area2\\ \hline
\multirow{3}{*}{\bf {PSS}}        &-0.568 $\pm$ 4.055i    & \bf{0.139}  & 0.645 & \bf{Inter-area}\\
{}                          &-1.983 $\pm$ 8.276i	&  0.233 & 1.317 & Intra-area1\\
{}                          &-3.011 $\pm$ 9.764i	&  0.295 & 1.554 & Intra-area2\\ \hline
\hline
\end{tabular}
\vspace{-0.20in}
\end{table}

\begin{table}[t]
\centering
\caption{Results of two-area four machine system with WF attached next to G4}
\label{tab:table2}
\renewcommand{\arraystretch}{1}
\begin{tabular}{||c|c|c|c|c||}
\hline \hline
\multirow{2}{*}{\parbox{1.1cm}{\centering {\bf Case 2: No PSS}\\}}  & 
\multirow{2}{*}{\parbox{1.4cm}{\centering {{{\bf Eigenvalue}}}}} & 
\multirow{2}{*}{\parbox{1.1cm}{\centering {{{\bf Damping}}}}} & 
\multirow{2}{*}{\parbox{0.8cm}{\centering {{{\bf Freq. \\(Hz)}}}}} & 
\multirow{2}{*}{\parbox{1.2cm}{\centering {{{\bf Oscillation \\ mode}}}}}  \\
{}       &      {}     &         {} &      {}     &       {} \\ \hline
\hline  
\multirow{3}{*}{\centering {10\%WG}}  & 0.094 $\pm$ 4.168i   & \bf{-0.023} & 0.663 & \bf{Inter-area}\\
{}                                          &-0.679 $\pm$ 7.048i   &  0.096 & 1.122 & Intra-area1\\
{}                                          &-0.625 $\pm$ 7.436i   &  0.084 & 1.183 & Intra-area2\\ \hline
\multirow{3}{*}{\centering {25\%WG}}    & 0.089 $\pm$ 4.320i   & \bf{-0.021} & 0.688 & \bf{Inter-area}\\ 
{}                                         &-0.680 $\pm$ 7.048i   &  0.096 & 1.122 & Intra-area1\\ 
{}                                          &-0.483 $\pm$ 7.720i   &  0.062 & 1.229 & Intra-area2\\ \hline
\multirow{3}{*}{\centering {35\%WG}}   & 0.084 $\pm$ 4.405i   & \bf{-0.019} & 0.701 & \bf{Inter-area}\\ 
{}                                          &-0.328 $\pm$ 7.934i   &  0.041 & 1.263 & Intra-area1\\ 
{}                                         &-0.680 $\pm$ 7.049i   &  0.096 & 1.122 & Intra-area2\\ \hline \hline
\multirow{2}{*}{\parbox{1.1cm}{\centering {\bf Case 2: PSS}\\}}  & 
\multirow{2}{*}{\parbox{1.4cm}{\centering {{{\bf Eigenvalue}}}}} & 
\multirow{2}{*}{\parbox{1.1cm}{\centering {{{\bf Damping}}}}} & 
\multirow{2}{*}{\parbox{0.8cm}{\centering {{{\bf Freq. \\(Hz)}}}}} & 
\multirow{2}{*}{\parbox{1.2cm}{\centering {{{\bf Oscillation \\ mode}}}}}  \\
{}       &      {}     &         {} &      {}     &       {} \\ \hline
\hline
\multirow{3}{*}{\centering {10\%WG}}    & -0.617	$\pm$ 4.221i&	\bf{0.145}&	0.672 & \bf{Inter-area}\\ 
{}                           &              -1.802	$\pm$ 8.394i&	0.210&	1.336 & Intra-area1\\
{}                            &             -3.011	$\pm$ 9.763i&	0.295&	1.554 & Intra-area2\\ \hline
\multirow{3}{*}{\centering {25\%WG}}   & -0.694	$\pm$ 4.410i&	\bf{0.155}&	0.702 & \bf{Inter-area}\\ 
{}                            &            -1.463	$\pm$ 8.577i&	0.168&	1.365& Intra-area1\\
{}                           &             -3.011	$\pm$ 9.761i&	0.295&	1.554& Intra-area2\\ \hline
\multirow{3}{*}{\centering {35\%WG}}   & -0.751	$\pm$ 4.522i&	\bf{0.164}&	0.720 & \bf{Inter-area}\\
{}                           &          -1.179	$\pm$ 8.703i&	0.134&	1.385& Intra-area1\\
{}                           &          -3.012	$\pm$ 9.760i&	0.295&	1.553& Intra-area2\\ \hline \hline
\end{tabular}
\vspace{-0.20in}
\end{table}

\begin{table}[t]
\centering
\caption{Results of two-area four machine system with WF attached next to L1}
\label{tab:table3}
\renewcommand{\arraystretch}{1}
\begin{tabular}{||c|c|c|c|c||}
\hline \hline
\multirow{2}{*}{\parbox{1.1cm}{\centering {\bf Case 3: No PSS}\\}}  & 
\multirow{2}{*}{\parbox{1.4cm}{\centering {{{\bf Eigenvalue}}}}} & 
\multirow{2}{*}{\parbox{1.1cm}{\centering {{{\bf Damping}}}}} & 
\multirow{2}{*}{\parbox{0.8cm}{\centering {{{\bf Freq. \\(Hz)}}}}} & 
\multirow{2}{*}{\parbox{1.2cm}{\centering {{{\bf Oscillation \\ mode}}}}}  \\
{}       &      {}     &         {} &      {}     &       {} \\ \hline
\hline 
\multirow{3}{*}{\centering {10\%WG}}    &0.141	$\pm$ 4.127i&	\bf{-0.034}&	0.657& \bf{Inter-area}\\
{}                                          &-0.604	$\pm$ 7.180i&	0.084&	1.143& Intra-area1\\
{}                                          &-0.665	$\pm$ 7.278i&	0.091&	1.158& Intra-area2\\ \hline
\multirow{3}{*}{\centering {25\%WG}}   &0.172	$\pm$ 4.156i&	\bf{-0.041}&	0.661& \bf{Inter-area}\\ 
{}                                          &-0.331	$\pm$ 7.104i&	0.047&	1.131& Intra-area1\\ 
{}                                         &-0.762	$\pm$ 7.115i&	0.106&	1.132& Intra-area2\\ \hline
\multirow{3}{*}{\centering {35\%WG}}    &0.192	$\pm$ 4.173i&	\bf{-0.046}&	0.664& \bf{Inter-area}\\ 
{}                                         &-0.131	$\pm$ 7.058i&	0.019&	1.123& Intra-area1\\ 
{}                                          &-0.834	$\pm$ 7.011i&	0.118&	1.116& Intra-area2\\ \hline \hline
\multirow{2}{*}{\parbox{1.1cm}{\centering {\bf Case 3: PSS}\\}}  & 
\multirow{2}{*}{\parbox{1.4cm}{\centering {{{\bf Eigenvalue}}}}} & 
\multirow{2}{*}{\parbox{1.1cm}{\centering {{{\bf Damping}}}}} & 
\multirow{2}{*}{\parbox{0.8cm}{\centering {{{\bf Freq. \\(Hz)}}}}} & 
\multirow{2}{*}{\parbox{1.2cm}{\centering {{{\bf Oscillation \\ mode}}}}}  \\
{}       &      {}     &         {} &      {}     &       {} \\ \hline
\hline
\multirow{3}{*}{\centering {10\%WG}}    &-0.603	$\pm$ 4.156i&	\bf{0.144}&	0.662& \bf{Inter-area}\\ 
{}                           &-1.984	$\pm$ 8.276i&	0.233&	1.317& Intra-area1\\
{}                            &-2.840	$\pm$ 10.057i&	0.272&	1.601& Intra-area2\\ \hline
\multirow{3}{*}{\centering {25\%WG}}    &-0.638	$\pm$ 4.172i&	\bf{0.151}&	0.664& \bf{Inter-area}\\ 
{}                            &-2.123	$\pm$ 8.176i&	0.251&	1.301& Intra-area1\\
{}                           &-2.633	$\pm$ 9.961i&	0.256&	1.585& Intra-area2\\ \hline
\multirow{3}{*}{\centering {35\%WG}}    &-0.660	$\pm$ 4.181i&	\bf{0.156}&	0.665& \bf{Inter-area}\\
{}                           &-2.221	$\pm$ 8.111i&	0.264&	1.291& Intra-area1\\
{}                         &-2.473	$\pm$ 9.893i&	0.243&	1.575&  Intra-area2\\ \hline
\hline
\end{tabular}
\vspace{-0.20in}
\end{table}

\begin{table}[t]
\centering
\caption{Results of two-area four machine system with WF attached next to L1 with controllers and 25\% Wind Generation}
\label{tab:table4}
\renewcommand{\arraystretch}{1}
\begin{tabular}{||c|c|c|c|c|c||}
\hline
\hline
\multirow{2}{*}{\parbox{1.1cm}{\centering {\bf Case 4}\\}}  & 
\multirow{2}{*}{\parbox{1.4cm}{\centering {{{\bf Eigenvalue}}}}} & 
\multirow{2}{*}{\parbox{1.1cm}{\centering {{{\bf Damping}}}}} & 
\multirow{2}{*}{\parbox{0.8cm}{\centering {{{\bf Freq. \\(Hz)}}}}} & 
\multirow{2}{*}{\parbox{1.2cm}{\centering {{{\bf Oscillation \\ mode}}}}}  \\
{}       &      {}     &          {} &      {}     &       {} \\ \hline
\hline  
\multirow{3}{*}{\bf {No PSS}}    &-0.060	$\pm$ 4.223i&	\bf{0.014}&	0.672& \bf{Inter-area}\\ 
{}                                          &-0.351	$\pm$ 7.029i&	0.050&	1.119& Intra-area1\\ 
{}                                          &-0.760	$\pm$ 7.117i&	0.106&	1.133& Intra-area2\\ \hline \hline
\multirow{3}{*}{\bf {PSS}}    &-0.933	$\pm$ 4.040i&	\bf{0.225}&	0.643& \bf{Inter-area}\\
{}                           &-2.121	$\pm$ 8.173i&	0.251&	1.301& Intra-area1\\
{}                          &-2.709	$\pm$ 9.828i&	0.266&	1.564& Intra-area2\\ \hline
\hline
\end{tabular}
\vspace{-0.20in}
\end{table}

From Tables \ref{tab:table1}, \ref{tab:table2}, \ref{tab:table3} and \ref{tab:table4} we observe that when the power system is integrated with WF, the damping of the system turns weaker without PSS control into the synchronous generators. However, when the PSS control is activated the inter-area  oscillations are damped. Comparing this observation when voltage and SDCs are added, the system becomes more stable and the damping is improved. Particularly, in cases 1 to 3 i.e. without integrating voltage controllers and SDCs, the inter-area oscillations have a negative damping contribution when PSS is deactivated. Nevertheless, in case 4, even without PSS activated, the controllers provide positive damping to the system at the inter-tie frequency. The comparison between the test cases when PSS is used shows a 62\% improvement when WF is attached next to L1 and all the controllers are switched-on. Even in the situation where voltage and damping controllers are deactivated, there is a 18\% improvement to the damping ratio with 35\% wind integration next to G4. In order to show the great effect of the controllers, we choose the comparison between the base system (Fig. \ref{f:fig6_label}) and the system presented in Fig. \ref{f:fig8_label} since the latter has slightly worse inter-area oscillation than the system shown in Fig. \ref{f:fig7_label}. The improvement by using the controllers is around 49\%. Finally, the overall improvement between the base study and the study where \begin{enumerate*}[label=\itshape\alph*\upshape)]
\item all the controllers are activated and 
\item 25\% wind generation is integrated to the system, 
\end{enumerate*} is around 933\% at 0.64Hz.

\section{Conclusions}\label{s:conclusions}

In this paper, the impact of DFIG based WF on the small-signal stability performance is investigated. For the conditions considered, the WF does not degrade the two-area system modal characteristics when the PSS controller is activated into the conventional machines. On the contrary, PSS control improves the damping of inter-area modes when WF penetration level is increasing. The results obtained also show that when the voltage and SDCs are added to the DFIG the system inter-area damping ratio improves dramatically.

\bibliographystyle{IEEEtran} 
\bibliography{ImpactWindOscillationsCK}

\end{document}